\documentclass[final]{appolb}
\usepackage{epsfig}
\newcommand{\eq}[1]{\begin{equation} #1 \end{equation}}

\begin{document}
\title{Critical fluctuations in models with van der Waals interactions%
\thanks{Presented at CPOD 2016, Wroclaw, Poland, May 30 -- June 4, 2016}%
}
\author{V.~Vovchenko$^{a,b,c}$, D.V.~Anchishkin$^{c,d}$, M.I.~Gorenstein$^{a,d}$, R.V.~Poberezhnyuk$^d$, H.~Stoecker$^{a,b,e}$
\address{
$^a$Frankfurt Institute for Advanced Studies, Goethe Universit\"at Frankfurt,
D-60438 Frankfurt am Main, Germany
\vskip2pt
$^b$Institut f\"ur Theoretische Physik,
Goethe Universit\"at Frankfurt, D-60438 Frankfurt am Main, Germany
\vskip2pt
$^c$Taras Shevchenko National University of Kiev, 03022 Kiev, Ukraine
\vskip2pt
$^d$Bogolyubov Institute for Theoretical Physics, 03680 Kiev, Ukraine
\vskip2pt
$^e$GSI Helmholtzzentrum f\"ur Schwerionenforschung GmbH, D-64291 Darmstadt, Germany
}
% \and
% the Name(s) of other Author(s)
% \address{and their affiliation}
}
\maketitle
\begin{abstract}
Particle number fluctuations are considered within the van der Waals (VDW) equation,
which contains both attractive 
(mean-field) 
and repulsive 
(eigenvolume) 
interactions.
The VDW equation is used to calculate the scaled variance
of particle number fluctuations in generic Boltzmann VDW system and
in nuclear matter.
The strongly intensive measures $\Delta[E^*,N]$ and $\Sigma[E^*,N]$ of the particle number and excitation energy fluctuations are also considered,
and, similarly, show singular behavior near the critical point. The $\Delta[E^*,N]$ measure is shown to attain both positive and negative values in the vicinity of critical point. Based on universality argument, similar behavior is expected to occur in the vicinity of the QCD critical point.
\end{abstract}
  
\section{Introduction}
The study of event-by-event  fluctuations in high-energy nucleus-nucleus
collisions is presently an essential task in the search of phase transitions and critical point (CP) in QCD~(see, e.g., Refs.~\cite{Koch:2008ia,fluc1} and references therein).
The van der Waals (VDW) equation of state is
a simple and popular
model to describe repulsive and attractive interactions
between particles~\cite{LL} and 
is suitable for studying the qualitative features of the critical behavior of fluctuations in the vicinity of the CP.
Two extensions of the classical VDW equation are considered:  (i) the grand canonical ensemble (GCE) formulation,  and (ii) the inclusion of the quantum statistics. 

\section{Extensions of the Van der Waals equation}

In the classical VDW equation the pressure is expressed as a function of temperature $T$, volume $V$, and particle number $N$ as
\eq{\label{vdw-p-n}
p(T,V,N)~=~\frac{NT}{V~-~b\,N}~-~a\,\frac{N^2}{V^2}~,
}
where $a$ and $b$ are, respectively, the attraction and repulsion VDW parameters.
Eq.~(\ref{vdw-p-n}) determines the pressure in the canonical ensemble (CE). 

However, the CE pressure
does not give a complete thermodynamical
description of the system. This is because parameters
$V$, $T$, and $N$ are  not the natural
variables for the pressure. The thermodynamical potential in the CE
is the free energy $F(T,V,N)$.
Recalling the thermodynamic identity $p = - (\partial F/ \partial V)_{T,N}$, and by requiring that for $a=0$ and $b=0$ the system reduces to the ideal gas, one obtains the free energy of the VDW gas
\eq{\label{vdw-F-n}
F(T,V,N)~=~F_{\rm id} (T,V-bN,N)~-~a\,\frac{N^2}{V}~,
}
where $F_{\rm id}$ is the free energy of the corresponding ideal gas.
The $F_{\rm id} (T,V,N)$ contains additional information which is missing in the standard VDW equation (\ref{vdw-p-n}) such as particle's mass $m$ and degeneracy $d$. The $F(T,V,N)$ function itself contains complete thermodynamic information about the system.

\subsection{Quantum statistics}
Relation (\ref{vdw-F-n}) is rigorously formulated for the case of Boltzmann statistics. In the following we postulate that it is also valid for quantum statistics, by assuming that $F_{\rm id} (T,V,N)$ is the free energy of the corresponding ideal \emph{quantum} gas. 
In such a way we obtain the VDW equation which includes the effects of quantum statistics.
The CE pressure reads
\eq{\label{vdw-p-n-q}
p(T,V,N) = - (\partial F/ \partial V)_{T,N} = p_{\rm id} (T, V - bN, N) - a \,\frac{N^2}{V^2}~,
}
where $p_{\rm id} (T, V, N)$ is the CE pressure of the ideal quantum gas.
For Boltzmann statistics $p_{\rm id} (T, V - bN, N) = NT / (V-bN)$, and in this case Eq.~(\ref{vdw-p-n-q}) coincides with Eq.~(\ref{vdw-p-n}).

The total entropy $S = -(\partial F/ \partial T)_{V,N}$ of the VDW gas reads $S(T,V,N) = S_{\rm id} (T, V-bN, N)$. One can easily see that entropy is always positive and that $S \to 0$ with $T \to 0$, in accordance with the 3rd law of thermodynamics.

\subsection{Grand canonical ensemble}
In the GCE the pressure $p(T,\mu)$ given as a function of its natural variables $T$ and $\mu$ contains complete information about the system. In order to transform the VDW equation to GCE we calculate the CE chemical potential as $\mu(T,V,N) = (\partial F/ \partial N)_{T,V}$, and, 
denoting $\mu \equiv \mu(T,V,N)$ and $N/V \equiv n$, solve resulting equation %(\ref{vdw-mu-n}) 
to obtain GCE particle density $n(T,\mu)$~\cite{VDWgen}
\eq{\label{vdw-n-gce}
n(T,\mu) = \frac{n_{\rm id}(T,\mu^*)}{1 + b\,n_{\rm id}(T,\mu^*)},
}
where $\mu^* = \mu - b\,p - a\,b\,n^2 + 2\,a\,n$.
The GCE pressure $p(T,\mu)$ reads $p(T,\mu) = p_{\rm id} (T, \mu^*) - a \, [n(T,\mu)]^2$.
At any given $T$ and $\mu$ equations for $p(T,\mu)$ and $n(T,\mu)$ should be solved simultaneously. It is possible that there is more than a single solution at a given $T$ and $\mu$. 
In that case the solution with the largest pressure should be chosen, in accordance with the Gibbs criteria.

\section{Calculation results}

\subsection{Particle number fluctuations}
The scaled variance of the total particle number fluctuations in the GCE can be
calculated as~\cite{VDWgen}
\eq{
\omega[N] = 
%\frac{k_2}{k_1} =
\frac{T}{n} \, \left(\frac{\partial n}{\partial \mu}\right)_T
= \omega_{\rm id} (T, \mu^*) \, \left[ \frac{1}{(1-bn)^2} - \frac{2an}{T} \,
\omega_{\rm id} (T, \mu^*) \right]^{-1},
\label{omegaN}
}
where $\omega_{\rm id}$ corresponds to the scaled variance of particle
number fluctuations in the ideal gas (in the Boltzmann approximation it is reduced to $\omega_{\rm id}=1$).

\begin{figure}[!t]
\centering
\includegraphics[width=0.49\textwidth]{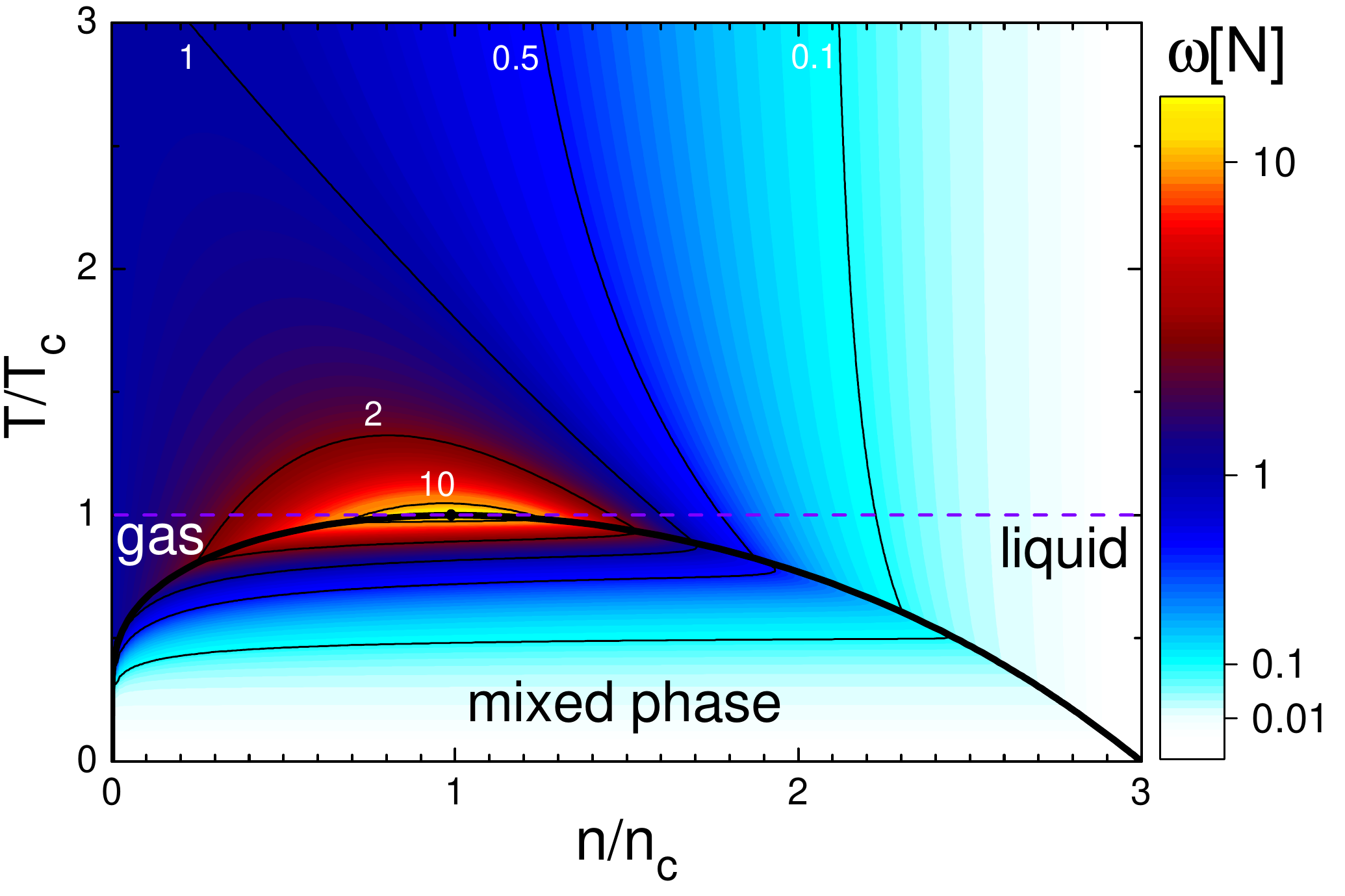}
\includegraphics[width=0.49\textwidth]{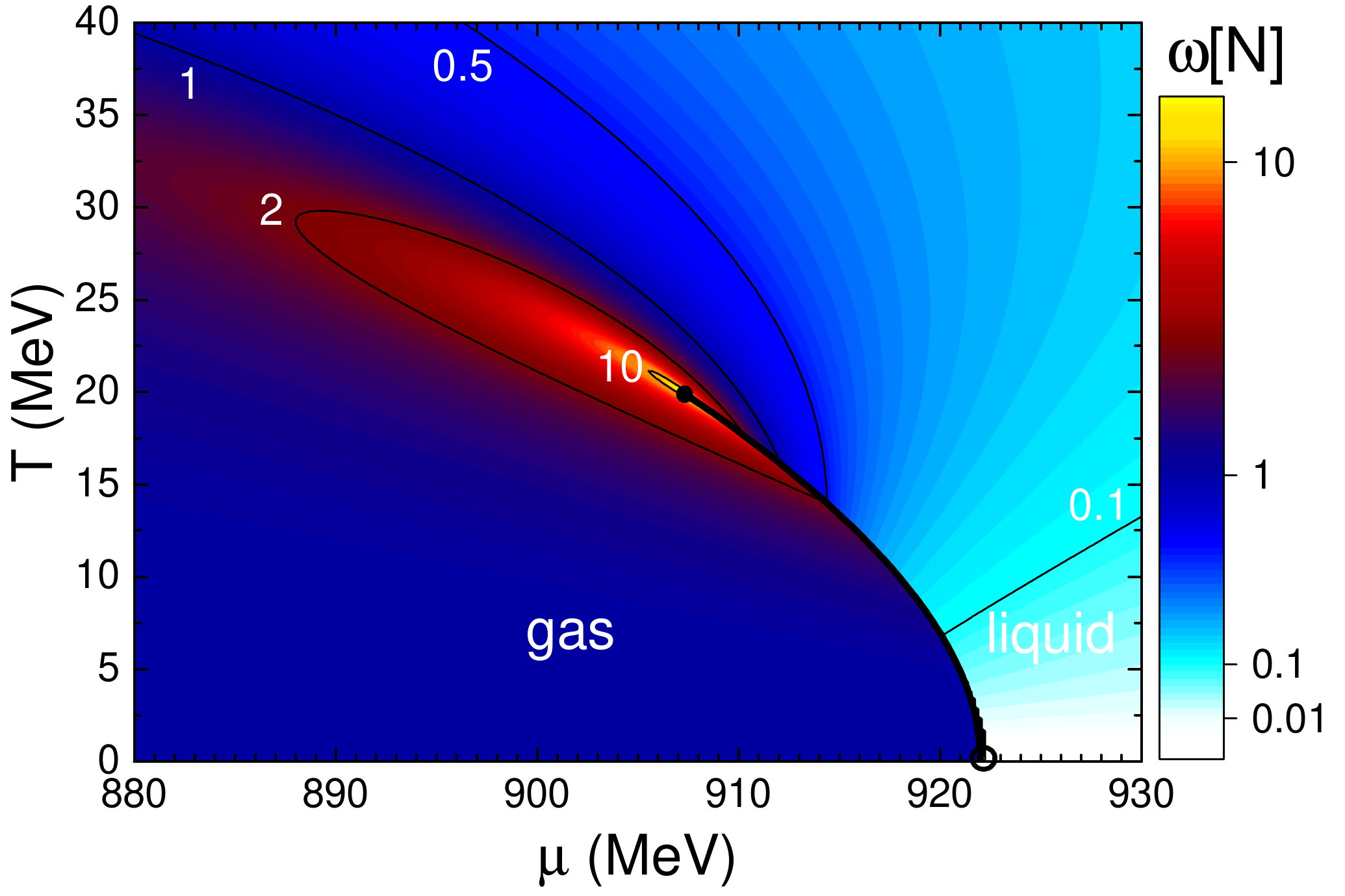}
\caption[]{
The density plots of scaled variance $\omega[N]$ of particle number fluctuations on the phase diagram of the classical VDW system (left panel) and of the nuclear matter (right panel). 
}\label{fig:wN}
\end{figure}

In our calculations we consider two representative VDW systems:
(1) the generic VDW system obeying Boltzmann statistics and (2) the
nuclear matter described as VDW equation with Fermi statistics for nucleons.
In the first case all quantities of interest do not explicitly depend on VDW parameters $a$ and $b$.
For the second case, following Ref.~\cite{VDWgen}, we take $a = 329$~MeV~fm$^3$ and $b = 3.42$~fm$^3$ for nucleons.
Results of the calculations of $\omega[N]$ for the Boltzmann VDW gas and for nuclear matter are shown in Fig.~\ref{fig:wN}. The $\omega[N]$ has a very similar qualitative behavior in both cases: it is close unity at small densities in the gaseous phase, close to zero at large densities in the liquid phase, and diverges at the CP.

\subsection{Strongly intensive quantities}
The results for $\omega[N]$ demonstrate
a strong increase of the particle number fluctuations in a vicinity
of the CP. 
However, the potential observation of these fluctuations can be masked by the fluctuations of the system
volume, which cannot be completely avoided in the heavy-ion collision experiments.

\begin{figure}[!t]
\centering
\includegraphics[width=0.49\textwidth]{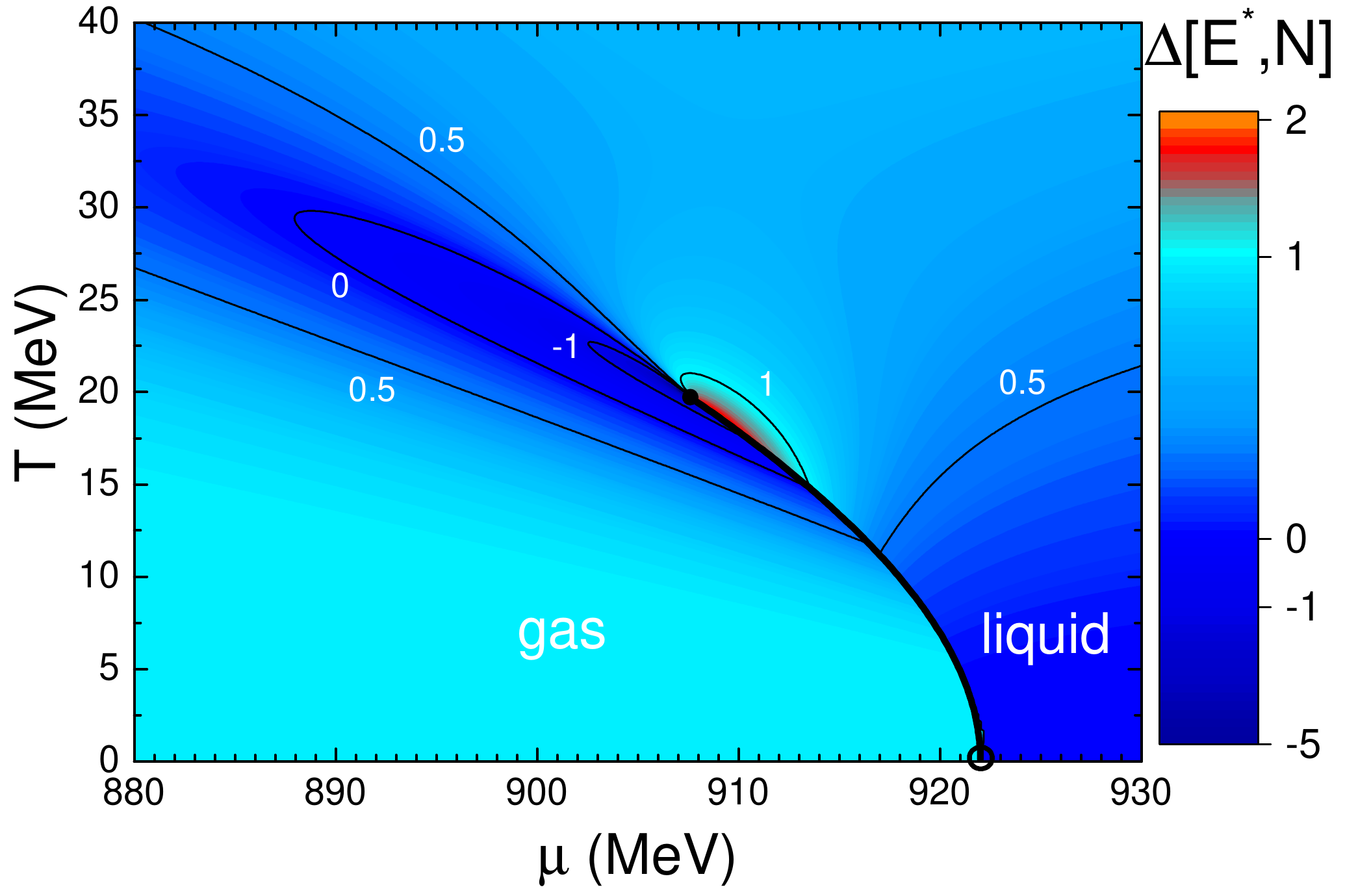}
\includegraphics[width=0.49\textwidth]{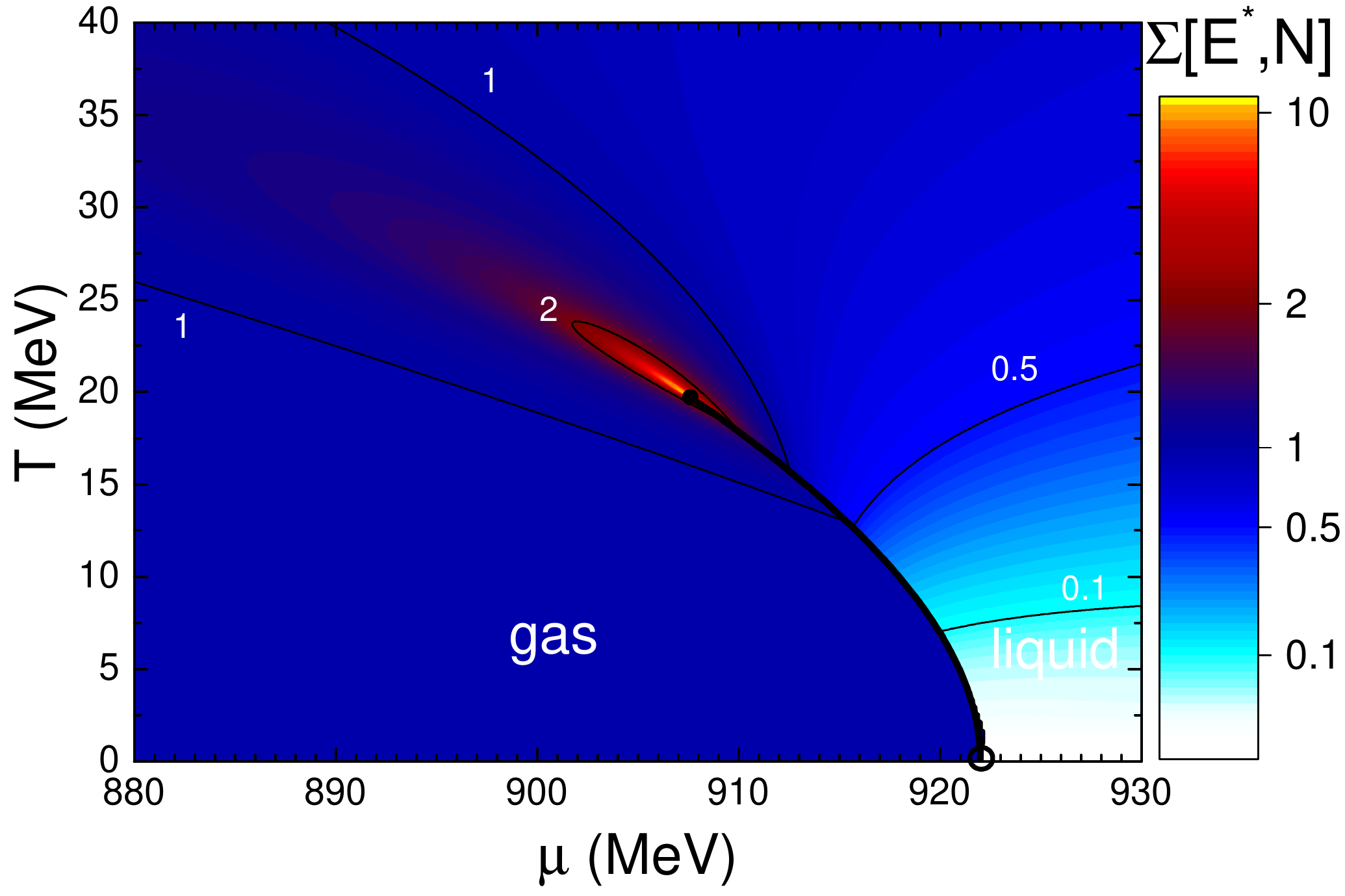}
\caption[]{
The density plots of the strongly intensive measures $\Delta[E^*,N]$ (left panel) and $\Sigma[E^*,N]$ (right panel) on the phase diagram of the nuclear matter. 
}\label{fig:SIQ}
\end{figure}

The strongly intensive measures
of the fluctuations defined in terms of two extensive quantities $A$ and $B$
were suggested in Ref.~\cite{SI}, and they are insensitive to the trivial fluctuations of the system volume.
Supposedly, these fluctuation measures also show critical behavior in the vicinity of CP. 
For example, they are used by the NA61/SHINE collaboration as probes for the QCD CP~
\cite{fluc1}. 
However, there is no proper model calculations which would confirm the presence of such critical behavior.
Thus, we consider the strongly intensive measures of the fluctuations of excitation energy $E^* = E - m_N\,N$ and nucleon number $N$ in VDW model.
The general expressions for these quantities read
\begin{eqnarray}
\Delta[E^*,N]~ &= & ~C_{\Delta}^{-1}\Big[\langle N \rangle \omega[E^*]-\langle E^* \rangle \omega[N] \Big]~,\\
\Sigma[E^*,N]~ &= & ~C_{\Sigma}^{-1}\Big[\langle N \rangle \omega[E^*]+\langle E^* \rangle \omega[N]
-2 \Big(\langle E^*N\rangle -\langle E^*\rangle\,\langle N\rangle\Big)\Big]~,\label{S}
\end{eqnarray}
where $\omega[E^*]$ is scaled variance of the excitation energy fluctuations, and
$C_{\Delta}^{-1}$ and $C_{\Sigma}^{-1}$ are the normalization factors
that have been suggested in the following form: $C_{\Delta} ~=~ C_{\Sigma} ~=~ \langle N \rangle \, \omega[\varepsilon^*]~$ with $\omega[\varepsilon^*]$ being the scaled variance of
a single-particle excitation (kinetic) energy distribution in the VDW system~\cite{SI}.
The details of the calculations of $\Delta[E^*,N]$ and $\Sigma[E^*,N]$ for classical VDW systems are described in Ref.~\cite{VDWgen}. 
For Fermi statistics the procedure is essentially the same. 
Results of calculations for the nuclear matter are shown in Fig.~\ref{fig:SIQ}.
Both the $\Delta[E^*,N]$ and $\Sigma[E^*,N]$ diverge at the CP, and, thus, appear to be suitable probes for the search of critical behavior. 
It is notable that $\Delta$ measure has a richer structure: for instance, it can take negative values in the vicinity of CP.

\section{Summary}
The critical behavior of different measures of fluctuations in systems with van der Waals interactions in the vicinity of the CP is demonstrated. In particular, model calculations confirm that the strongly intensive measures of energy and particle number fluctuations are
suitable probes of the critical behavior. 
The VDW interactions should also play a significant role in the full hadron resonance gas, in the regions of phase diagram away from nuclear matter, for instance where the chemical freeze-out in heavy-ion collisions can be expected to occur. This has recently been demonstrated for the crossover region at $\mu_B=0$ in Ref.~\cite{VDWHRG}. Further studies in this direction can shed new light on the role of VDW interactions in the confined phase of QCD.

\end{document}